\documentclass[journal]{IEEEtran}
\IEEEoverridecommandlockouts
\usepackage{cite}
\usepackage{amsmath,amssymb,amsfonts}
\usepackage{algorithmic}
\usepackage{graphicx}
\usepackage{textcomp}
\usepackage{xcolor}
\usepackage[inline]{enumitem}
\usepackage{float}
\usepackage{booktabs}
\usepackage{multirow}
\usepackage{tabularx}
\usepackage{subfigure}
\def\BibTeX{{\rm B\kern-.05em{\sc i\kern-.025em b}\kern-.08em
    T\kern-.1667em\lower.7ex\hbox{E}\kern-.125emX}}
\begin{document}

\title{Hy-DAT: A Tool to Address Hydropower Modeling Gaps Using Interdependency, Efficiency Curves, and Unit Dispatch Models
\thanks{Pacific Northwest National Laboratory is operated for DOE by Battelle Memorial Institute under Contract DE-AC05-76RL01830. This work was supported by DOE's Water Power Technology's Office.} 
\thanks{Corresponding author e-mail: sohom.datta@pnnl.gov.}

}

\author{
\IEEEauthorblockN{Dewei Wang, Bhaskar Mitra, Sameer Nekkalapu, Sohom Datta, Bibi Matthew, Rounak Meyur, Heng Wang, and Slaven Kincic}\\
\IEEEauthorblockA{Pacific Northwest National Laboratory, Richland, WA, USA}
}

\maketitle
\vspace{-1em}
\begin{abstract}
As the power system continues to be flooded with 
intermittent resources, it becomes more important to 
accurately assess the role of hydro and its impact on the power 
grid. While hydropower generation has been studied for decades, 
dependency of power generation on water availability and 
constraints in hydro operation are not well represented in 
power system models used in the planning and operation of large-scale interconnection studies. There are still multiple modeling 
gaps that need to be addressed; if not, they can lead to 
inaccurate operation and planning reliability studies, and 
consequently to unintentional load shedding or even 
blackouts. As a result, it is very important that
hydropower is represented correctly in both steady-state and
dynamic power system studies. In this paper, we discuss the  development and use of the Hydrological Dispatch and Analysis Tool (Hy-DAT) as an interactive graphical user interface, that uses a novel methodology to address the hydropower modeling gaps like water availability and interdependency using a database and algorithms to generate accurate representative models for power system simulation.

\end{abstract}




\section{Introduction}
 In the evolving power system, where new renewable 
resources displace continually conventional generation and hydro 
becomes a single traditional resource that is fully controllable, it 
is timely to examine hydropower generation’s role in the power system 
of the future and its representation in power system operation 
and planning reliability studies. Energy generation through hydroelectricity has a vital role in the development of a region and its economy. Hydropower generation depends on a lot of external factors such as water availability due to annual precipitation and snow water melt. As a result, representing hydropower generation accurately in electrical power system models requires proper translation of hydrology into the electrical capability of the plants. As renewable generation penetration continues to increase, hydropower operation has evolved from a baseload resource to a regulating resource; therefore,  it is very important that hydropower is represented correctly in both steady-state and dynamic power system studies \cite{mitra2023gaps}. These gaps in hydropower modeling are mainly a consequence of decoupling hydro conditions and hydro-based constraints with electrical models used in steady-state and dynamic studies \cite{kincic2023hydropower, kosterev}. The 
following gaps have been identified in close collaboration with the industry:
\begin{itemize}
  \item Water availability in steady-state and dynamic models are not properly represented. 
\item Interdependencies among hydro projects in the same river systems are not properly represented. 
\item Environmental constraints are not represented in models. 
\item Rough zones are not represented in the power system 
model so generation dispatch in system studies might 
not be realistic. 
\item Many outdated dynamic models of hydropower generation 
turbines are still in use. 
\end{itemize}
In order to conduct more accurate planning and operation 
studies, and ensure the grid operates reliably, it is vital to more accurately model hydropower generation. In this paper, we present the development of a tool that helps to address the hydropower modeling gaps with respect to water availability and interdependencies through the usage of databases and specific tools to translate historical hydrology into representative models that can be used in major power system simulations. The paper has the following sections: Section \ref{sec:db} provides the database components and features essential for the tool development. Section \ref{sec:eff} provides insight into application of the database for various tools like the  efficiency curve estimation, interdependency evaluation using historical datasets, and regression mode used for unit-level dispatch of hydropower resources to be integrated with Hy-DAT. Section \ref{sec:inter} provides integration of the various features for power system model updates and a reference case study. Finally, Section \ref{sec:con} concludes the paper with some discussions and future work.

\vspace{-1.5em}
\section{Database Generation and Utilization} \label{sec:db} 
One of the major objectives of this work is to couple the historical hydro data available in the literature \cite{USACE}, \cite{Britishhydro} with the electrical data. The unit-level generation dispatch data obtained from the Supervisory Control and Acquisition Data (SCADA) measurements are taken as an energy management system snapshot of the system operating condition that can be used in the various planning cases considered in this work. This hydro data and the electrical data for various plants and units has been uploaded to the Database Management System DB Browser \cite{DB} software tool using the structured query language (SQL) libraries through various python scripts developed as part of this work. 
    
\subsection{Source, Structure, and Data Description}
\label{sub:2a}
The timeseries hydro data and unit-level MW dispatch data for this work have been populated into the database and divided into six sections whose details are presented below:

\begin{itemize}
 \item \textbf{Plant\textunderscore Data}: The hydro data in this work is in timeseries comprising 10 years of data (2013-2023), with a sampling frequency of 1 hour, for various hydro plants across the Columbia River Basin \cite{USACE}. 

    \item \textbf{Unit\textunderscore Data}: This table has been populated with the electrical data obtained using the Transient Security Assessment Tool (TSAT) \cite{TSAT}  to convert the energy management system data obtained from SCADA to capture the unit-level dispatch data. 

    \item \textbf{Static\textunderscore Plant\textunderscore Data}: This table contains plant-level data, obtained from a 2020 Western Electricity Coordinating Council summer planning case and \cite{USACEMIL}. The table has location-based data such as latitude, longitude, and area number of the plant and generic hydro-based data such as head value of the plant. 

    \item \textbf{Static\textunderscore Unit\textunderscore Data}: Similar to the \textit{Static\textunderscore Plant\textunderscore Data} table, this table contains unit-level static data. 

    \item \textbf{Efficiency\textunderscore Raw\textunderscore Data}: This table contains the calculated efficiencies for each individual generation unit using the flow data from the \textit{Plant\textunderscore Data} table. 

    \item \textbf{Efficiency\textunderscore Estimated\textunderscore Data}: Hydro data for the plants are not available throughout the operating range of a generation unit to account for electrical power output; therefore, a linear regression model has been developed to extrapolate the missing estimated flow and estimated electrical power values for the individual units. 
    
\end{itemize}
\vspace{-1em}

\subsection{Database Schema and Database Instances}
\label{sub:2b}
The database schema developed in this work has been linked together using a common identifier as shown in Table \ref{Table 1}. This method of linking the tables using a common column identifier becomes important during the graphic user interface (GUI) development process of this work as discussed in Section \ref{integration} of this work.

\begin{table}[]
\centering
\caption{Subset Classification of Tables Based on the Considered Common Column Identifiers}
\label{Table 1}
\begin{tabular}{|c|c|c|}
\hline
\textbf{Subset} & \textbf{\begin{tabular}[c]{@{}c@{}}Column Identifier\\ Name\end{tabular}} &  \textbf{Tables} \\ \hline
1 & Project\_Name &  \begin{tabular}[c]{@{}c@{}}Plant\_Data,\\ Unit\_Data,\\ Static\_Unit\_Data,\\ Static\_Plant\_Data\end{tabular} \\ \hline
2 & Unit\_ID & \begin{tabular}[c]{@{}c@{}}Efficiency\_Raw,\\ Efficiency\_Estimated\end{tabular} \\ \hline
\end{tabular}
\end{table}

In Table \ref{Table 1}, it should be noted that the tables in Subset 2 are linked to the tables in Subset 1 through the \textit{Unit\textunderscore ID} data available in the \textit{Unit\textunderscore Data} table. The \textit{Unit\textunderscore ID} of any unit in the database is populated as a string concatenation of the individual unit’s bus number and \textit{ID} (unit identifier name) available from the \textit{Unit\textunderscore Data} table. The \textit{Project\textunderscore Name} column identifier is also used to link the tables within Subset 1.

The database instance is a single snapshot of the database containing the data, where a single instance representing the \textit{`tatic\textunderscore Unit\textunderscore Data} table has been presented in Table \ref{Table 2}.

\begin{table}[ht!]
\centering
\caption{Database Instance of Static\textunderscore Unit\textunderscore Data Table}
\label{Table 2}

\begin{tabular}{|c|c|c|c|c|c|c|c|}
\hline
\textbf{\begin{tabular}[c]{@{}c@{}}Project\\ Name\end{tabular}} & \textbf{\begin{tabular}[c]{@{}c@{}}Bus\\ Name\end{tabular}} & \textbf{\begin{tabular}[c]{@{}c@{}}Bus\\ \#\end{tabular}} & \textbf{ID} & \textbf{\begin{tabular}[c]{@{}c@{}}Unit \\ ID\end{tabular}} & \textbf{\begin{tabular}[c]{@{}c@{}}Nom.\\ $P_{max}$\\ MW\end{tabular}} & \textbf{\begin{tabular}[c]{@{}c@{}}SCADA\\Bus\\\#\end{tabular}} & \textbf{\begin{tabular}[c]{@{}c@{}}SCADA\\BUS\\ID\end{tabular}} \\ \hline
\begin{tabular}[c]{@{}c@{}}Plant\\A \end{tabular} & \begin{tabular}[c]{@{}c@{}}Plant\\A\\ Bus1\end{tabular} & \begin{tabular}[c]{@{}c@{}}Plant\\A\\ Bus\\ \#1\end{tabular} & \begin{tabular}[c]{@{}c@{}}A \\ ID1\end{tabular} & \begin{tabular}[c]{@{}c@{}}Plant\\A\\ Bus\#\\1-A \\ ID1\end{tabular} & $X_1$ & \begin{tabular}[c]{@{}c@{}}Plant\\A \\Bus\#1\end{tabular} & \begin{tabular}[c]{@{}c@{}}A \\ ID1\end{tabular} \\ \hline
\begin{tabular}[c]{@{}c@{}}Plant\\A \end{tabular}& \begin{tabular}[c]{@{}c@{}}Plant\\A\\ Bus1\end{tabular} & \begin{tabular}[c]{@{}c@{}}Plant\\A\\ Bus\\ \#1\end{tabular} & \begin{tabular}[c]{@{}c@{}}A\\ID2\end{tabular} & \begin{tabular}[c]{@{}c@{}}Plant\\A   \\ Bus\#\\1-A \\ ID2\end{tabular} & $X_2$ & \begin{tabular}[c]{@{}c@{}}Plant\\A \\ Bus\#1\end{tabular} & \begin{tabular}[c]{@{}c@{}}A \\ ID2\end{tabular} \\ \hline
\begin{tabular}[c]{@{}c@{}}Plant\\A \end{tabular}& \begin{tabular}[c]{@{}c@{}}Plant\\A \\ Bus2\end{tabular} & \begin{tabular}[c]{@{}c@{}}Plant\\A\\ Bus\\\#2\end{tabular} & \begin{tabular}[c]{@{}c@{}}A \\ ID3\end{tabular} & \begin{tabular}[c]{@{}c@{}}Plant\\A \\ Bus\#\\2-A \\ ID3\end{tabular} & $X_3$ & \begin{tabular}[c]{@{}c@{}}Plant\\A \\ Bus\#2\end{tabular} & \begin{tabular}[c]{@{}c@{}}A \\ ID3\end{tabular} \\ \hline
\begin{tabular}[c]{@{}c@{}}Plant\\B \end{tabular}& \begin{tabular}[c]{@{}c@{}}Plant\\B \\ Bus1\end{tabular} & \begin{tabular}[c]{@{}c@{}}Plant\\B\\ Bus\\ \#1\end{tabular} & \begin{tabular}[c]{@{}c@{}}B\\ ID1\end{tabular} & \begin{tabular}[c]{@{}c@{}}Plant\\B   \\ Bus\#\\1-A \\ ID1\end{tabular} & $Y_1$ & \begin{tabular}[c]{@{}c@{}}Plant\\ B \\ Bus\#1\end{tabular} & - \\ \hline
\end{tabular}
\end{table}

\vspace{-1.5em}
\section{Application of Database for various tools}
\label{sec:eff}
\subsection{Efficiency Curve Evaluation}
Turbines are a vital component for hydropower plants, they are responsible for conversion of the kinetic energy into mechanical energy that in turn rotates the generators to produce electricity. 
Lack of proper estimation of turbine efficiency often leads to underestimating or overestimating the annual energy production capacity \cite{BelgiumHydro}. In most scenarios efficiency is considered as a static number and leads to erroneous estimates. Efficiency is a function of flow and varies with the flow rate through a turbine, a representative efficiency curve for different turbines has been discussed in \cite{Britishhydro}. Data-driven methods have been utilized to approximate efficiency curves for different plants; however, availability of granular data for different plants is difficult. A generic approach cannot be used to estimate the efficiency curve for another plant. 

It is preferred to operate hydropower turbines around the 90\% efficiency range to minimize the effect of stress on the operating turbines. Several strategies involving addition of battery energy storage to reduce such scenarios have been discussed in \cite{BHATTI2023120894}; however, the gap of estimating hydropower efficiency for particular flow regions does exist and are essential to compute the power dispatch conditions. 

\begin{figure}
    \centering
    \includegraphics[width=0.8\linewidth]{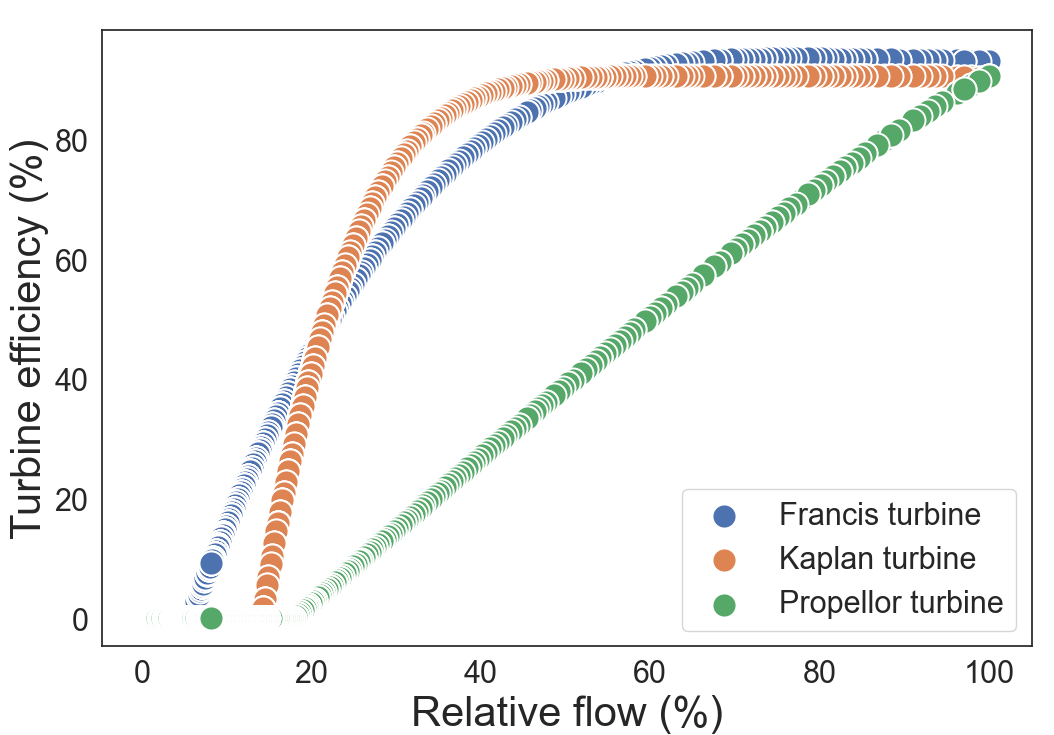}
    \caption{Efficiency curves for different turbines using HydroGenerate.}
    \label{fig:hydrogenerate_ex}
\end{figure}

To estimate efficiency curves for different turbines and different head conditions we utilized an open-source tool called HydroGenerate \cite{Bhaskar_hydrogenerate}. It is a web-based tool that estimates the power generation of a powered or nonpowered dam  through some simple inputs. It offers a wide selection of turbine types for selection. The detailed equations essential to calculate the efficiency curve has been discussed in \cite{mitra2020development}. An example use of the tool determining efficiency points for different relative flows of various turbines is shown in Fig. \ref{fig:hydrogenerate_ex}. For this analysis head was considered constant; however, in reality head varies with season, flow conditions, and several other factors. With variation of head, the operating efficiency points vary that would result in inaccurate estimates of power dispatch.

\begin{figure}[]
     \centering
     \subfigure[]{\includegraphics[width=0.8\linewidth]{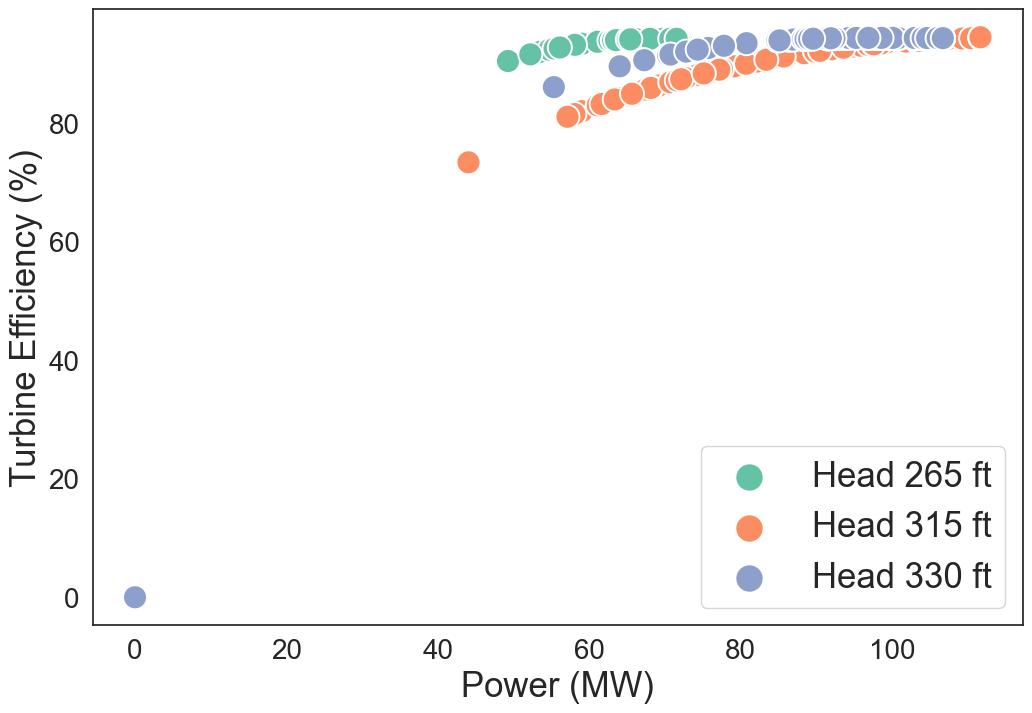}} 
      \vspace{-3mm}
     \subfigure[]{\includegraphics[width=0.8\linewidth]{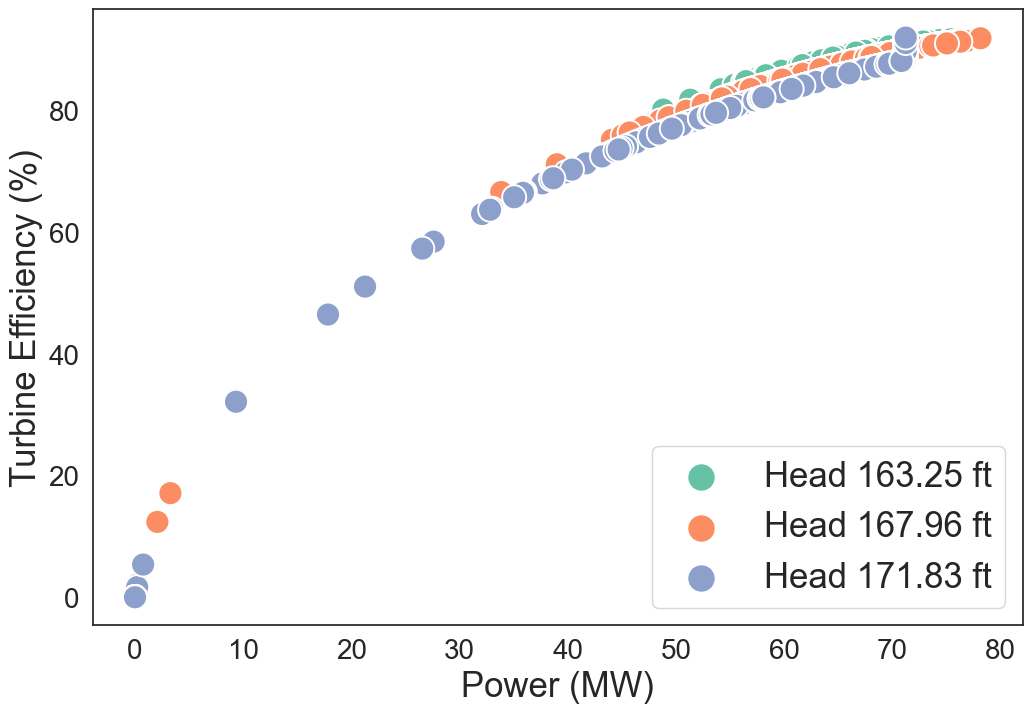}}
      \vspace{-0.7mm}
      \caption{Operational efficiency of a Francis turbine for two plants located in the western United States. (a) Plant A with three operating heads 265 ft., 315 ft., and 330 ft. and (b) Plant B with three operating heads 163.25 ft., 167.96 ft., and 171.83 ft.}
    \label{fig:comp}
\end{figure}

By using data from two different plants located in the continental western United States we calculate the efficiency parameters as shown in Fig. \ref{fig:comp}(a) and (b). 

\section{Interdependency Evaluation} \label{sec:inter}

Several factors contribute to how hydropower plants located in different watersheds respond. Internal response is affected by the operation of the storage reservoirs upstream by controlling the volume of water stored or released. There are several external factors like minimum water limits, fish habitats, recreation, etc. that restrict its operation. There is a mix of complex rules, policies, and agreements from the water management authorities responsible for maintaining water regulations that are regulated by Federal laws. There are three major watersheds in the continental United States: the Columbia River, the Colorado River, and the Tennessee River. 

The Columbia River Basin has its largest foothold in the Pacific Northwest Region having 250 reservoirs and 150 hydroelectric projects \cite{columbia}. Streamflow in the Columbia River watershed does not follow the electricity demand, where the peak is mostly observed during the winter months; however, the flows peak in spring and early summer when the snow pack melts, although recent changes in climate conditions have made some of those patterns erratic. The high rate of correlation between power generation can be used to understand the operating range for patterns in the plants located in the Columbia River Basin. The variation is not limited by flow but is also a function of season and head.
\vspace{-1em}

\subsection{Methodology}

\begin{figure}
    \centering
    \includegraphics[width=0.8\linewidth]{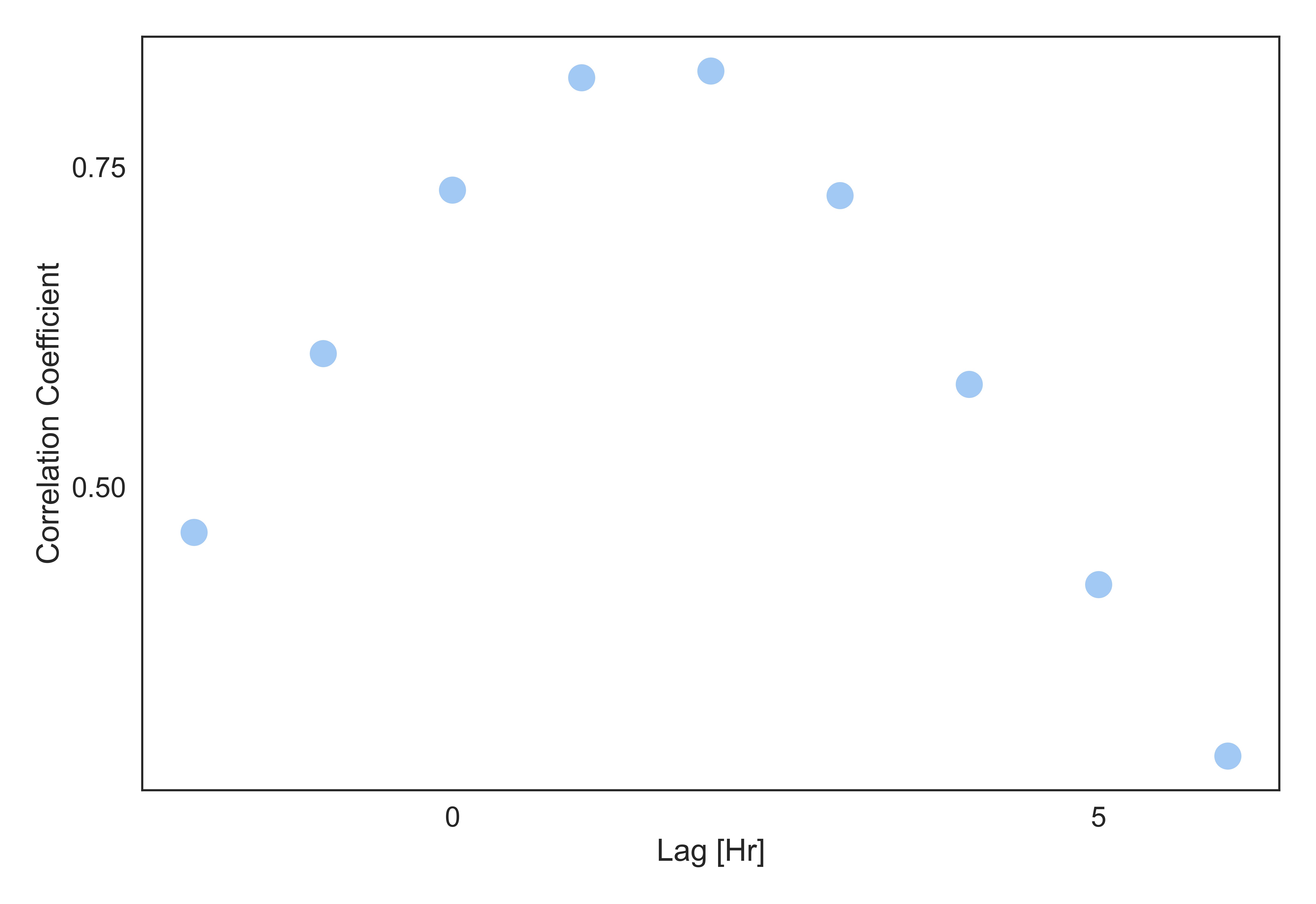}
    \caption{Time correlation of streamflow between Plant A and B during Winter.}
     \vspace{-5mm}
    \label{fig:corr_winter}
\end{figure}

Considering water time constant and distance between the two plants it is essential to establish a time correlation between the streamflow data. Data were split into three prominent seasons: \textit{Winter}, \textit{Spring}, and \textit{Summer}. As discussed before, streamflow depends on the season and thus the time lag for each season needs to be evaluated to establish a comprehensive understanding. Upon analysis, the time correlation for the three seasons were seen to peak, the highest between 1 and 2 hours. A representative plot for \textit{Winter} is shown in Fig. \ref{fig:corr_winter}.

Using this information, the original dataset was reorganized to match the time delay to establish a relationship between their power dispatch. For the analysis we further incorporated the head parameter of the upstream plant that influences the power generation downstream.

\begin{figure}[]
     \centering
     \subfigure[]{\includegraphics[width=0.8\linewidth]{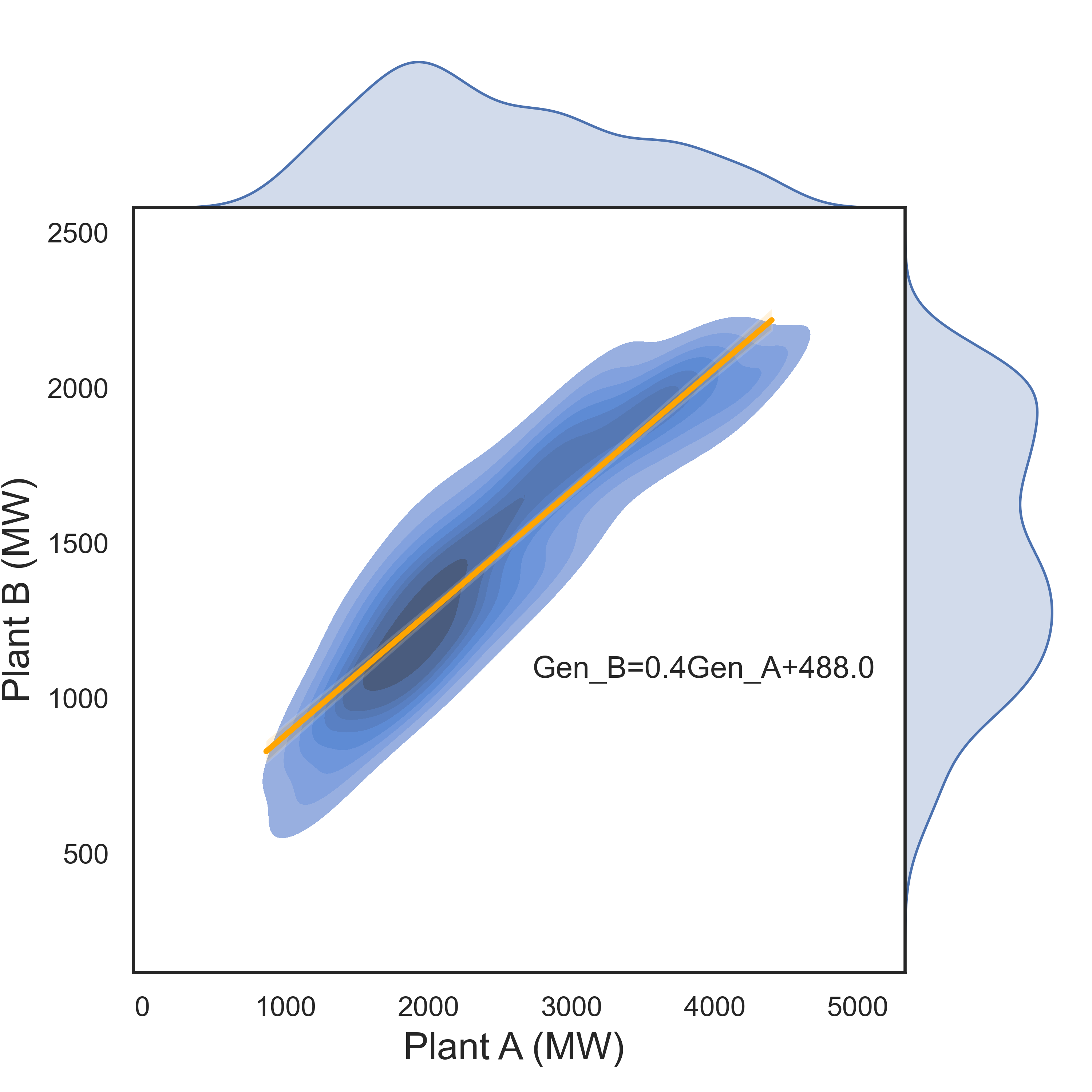}} 
     \vspace{-3mm}
     \subfigure[]{\includegraphics[width=0.8\linewidth]{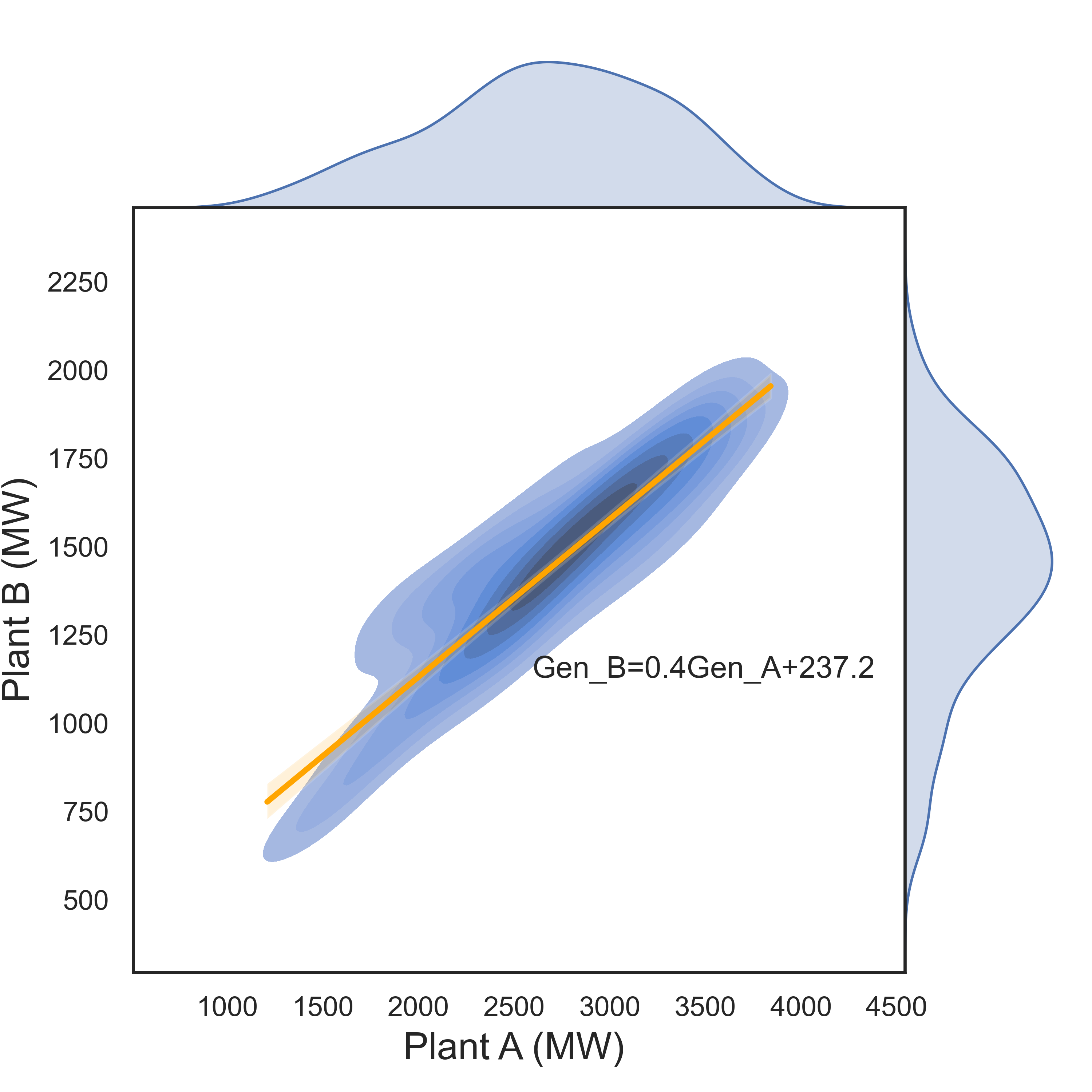}}
     \vspace{-0.9mm}
     \caption{Generation comparison for Winter. (a) Head for upstream plant (Plant A) is $294.34 ft.$ (b) Head for upstream plant (Plant A) is $322.45 ft.$}
    \label{fig:den_comp}
    
\end{figure}

As seen in Fig. \ref{fig:den_comp},  within the same season the generation for the two plants A and B show a significant variation with the change in head. A similar significance can be seen for the \textit{Summer} season when variation of head is considered.
A relationship between the generation patterns of the rivers upstream and downstream were obtained using linear regression of the power dispatch.
\vspace{-1em}
\subsection{Unit-level Dispatch using Regression Models}
With plant-level time series recordings and unit-level measurements, one can build regression models to correlate water head, total power output, etc., with individual generator status and power output. In this study, machine learning, more specifically deep neural networks (DNNs), is implemented to build regression models in view of its rapid development and wide applications in many engineering problems. Previous studies \cite {wang2021machine} have proven the high precision of DNN-based regression models and the impressive advantages over the conventional regression model construction approaches. A six-layer, fully connected DNN has been utilized for this approach.

\begin{figure}
    \centering
    \includegraphics[width=0.8\linewidth]{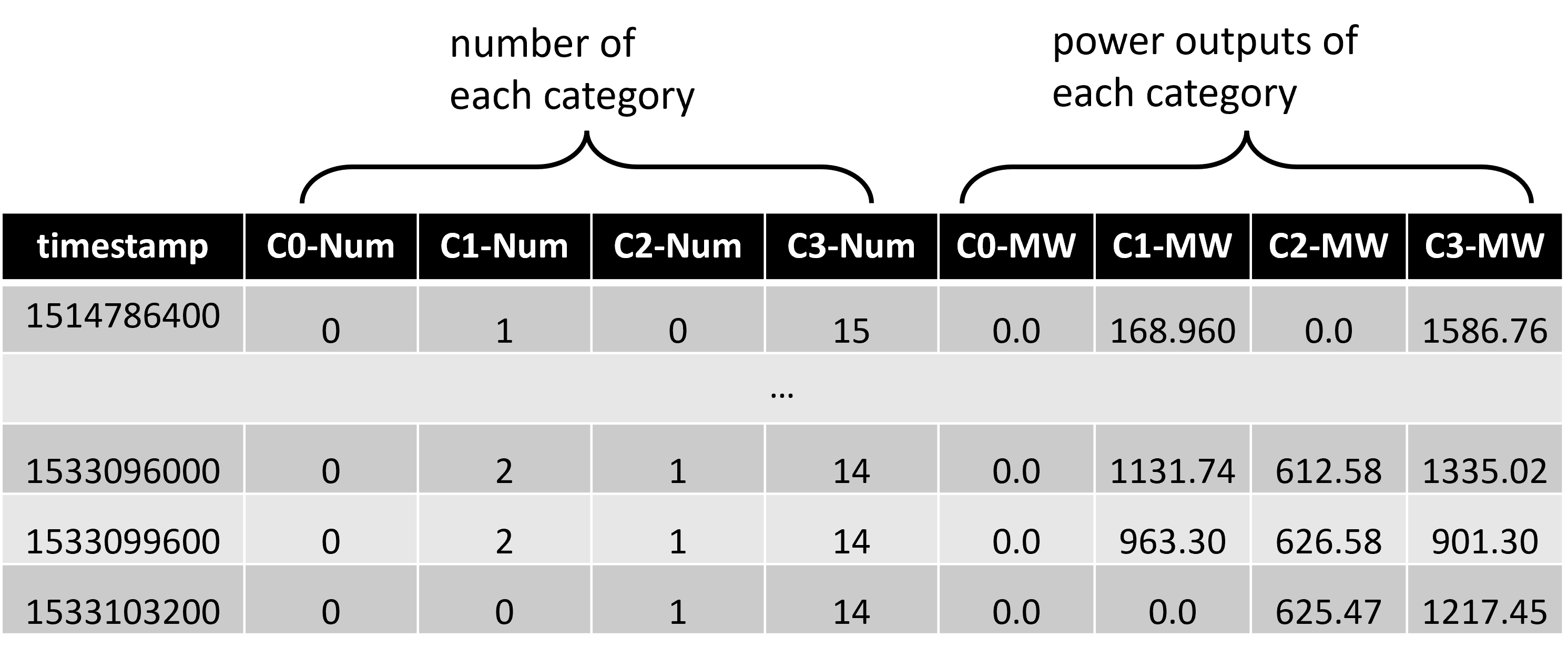}
    \caption{Unit-level data as model outputs.}
     \vspace{-3mm}
    \label{fig:ModelOutput}
\end{figure}

For upstream plant A, there are a total of 5,089 recordings at each hour in three seasons. Each recording comprises plant-level and unit-level data. Plant-level data contain total power output, waterhead, water storage, generator flow rate, spillage, etc. Power output, waterhead, and water storage were selected as model input parameters. Unit-level data contain the status and power outputs of 30 generators. These generator units can be classified into four categories (C1, C2, C3, and C4) according to nominal power. Status and power outputs of individual generators can be converted into a number of active units and power output of the four categories, as shown in Fig. \ref{fig:ModelOutput}.

\subsubsection{Training}

\begin{figure}[h]
    \centering
    \includegraphics[width=0.9\linewidth]{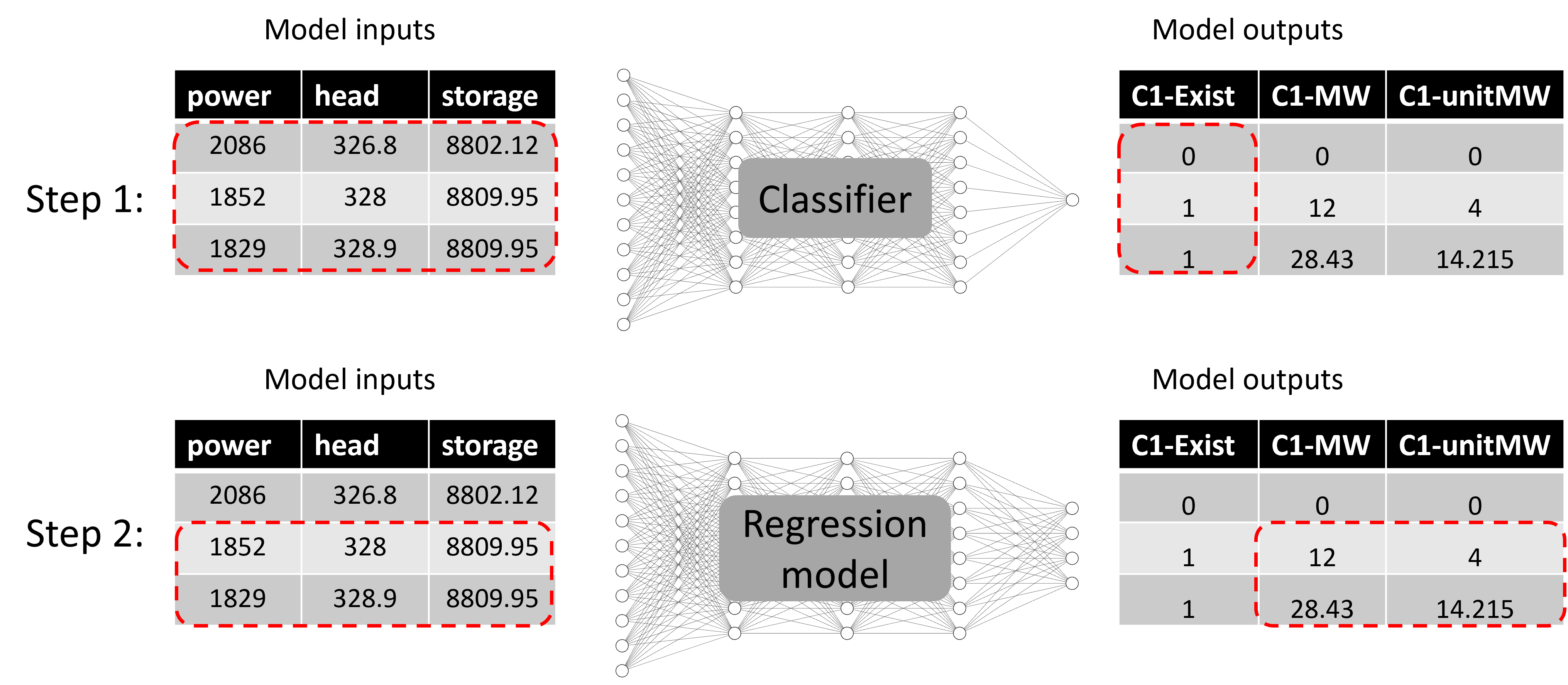}
    \caption{Model training procedure.}
     \vspace{-3mm}
    \label{fig:Train}
\end{figure}

Fig. \ref{fig:Train} shows the two-step model training procedure. As can be seen in the right plots, the number of active units and the power output of C1 were reorganized as three columns: C1-Exist indicates if there is an active unit in this category, C1-MW is the total power of this category, and C1-unitMW is the average power output of each active unit in this category. In Step 1, a DNN-based classifier is trained to correlate model inputs with C1-Exist. In Step 2, a DNN-based regression model is developed between model inputs and C1-MW, C1-unitMW for active data only (data in red dotted line). For the models in Fig. \ref{fig:Train}, the DNN-based classifier can predict whether C1 has active units with an accuracy of 91.45\%. The 80\% confidence intervals of the regression model are smaller than 10\% of the mean values.

\subsubsection{Testing and Results}
\begin{figure}[h]
    \centering
    \includegraphics[width=0.9\linewidth]{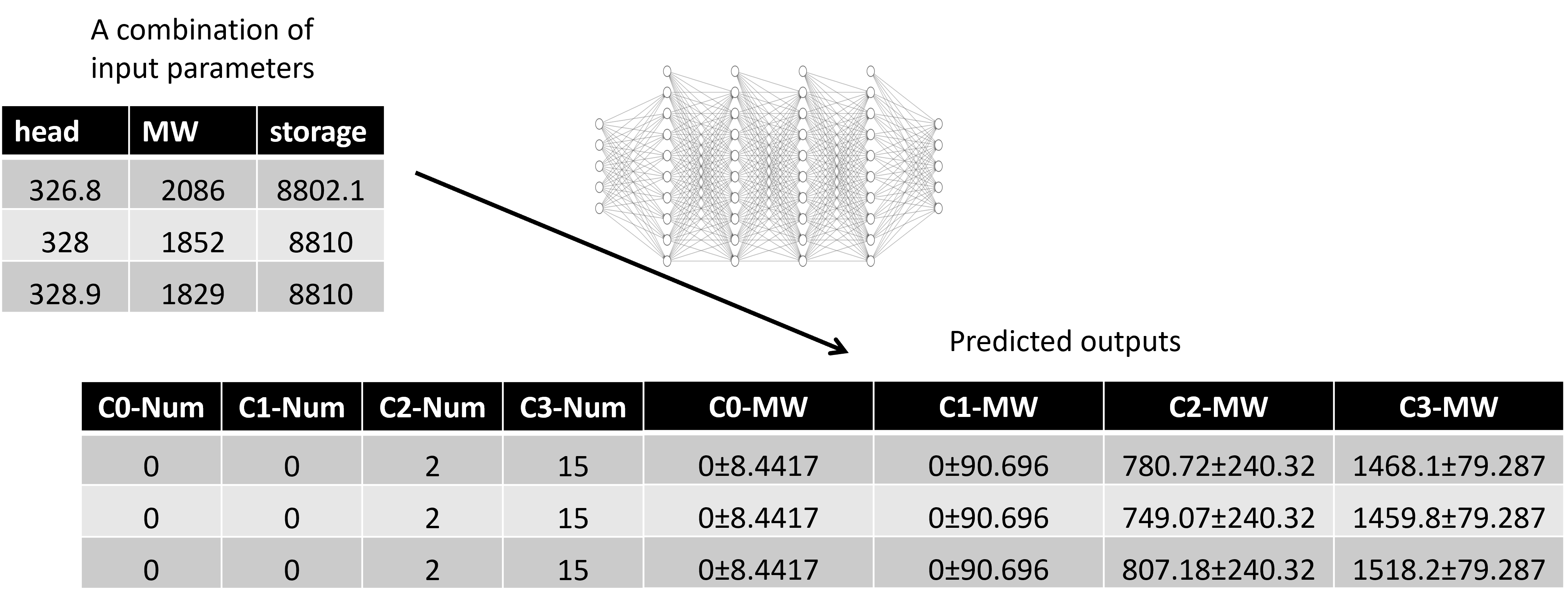}
    \caption{An example of model prediction for plant A.}
    \label{fig:PredictExample}
\end{figure}

By repeating the above procedure, one can obtain regression models for all categories. With these trained models, one can predict unit-level conditions from an appropriate combination of plant-level parameters. An example is given in Fig. \ref{fig:PredictExample}. Later, by allocating the total power output to the active units in each category, unit-level dispatch can be conducted according to the plant-level inputs.

\vspace{-1.5em}
\subsection{Integration and GUI Development} \label{integration}
\begin{figure}
    \centering
     \includegraphics[width=0.9\linewidth]{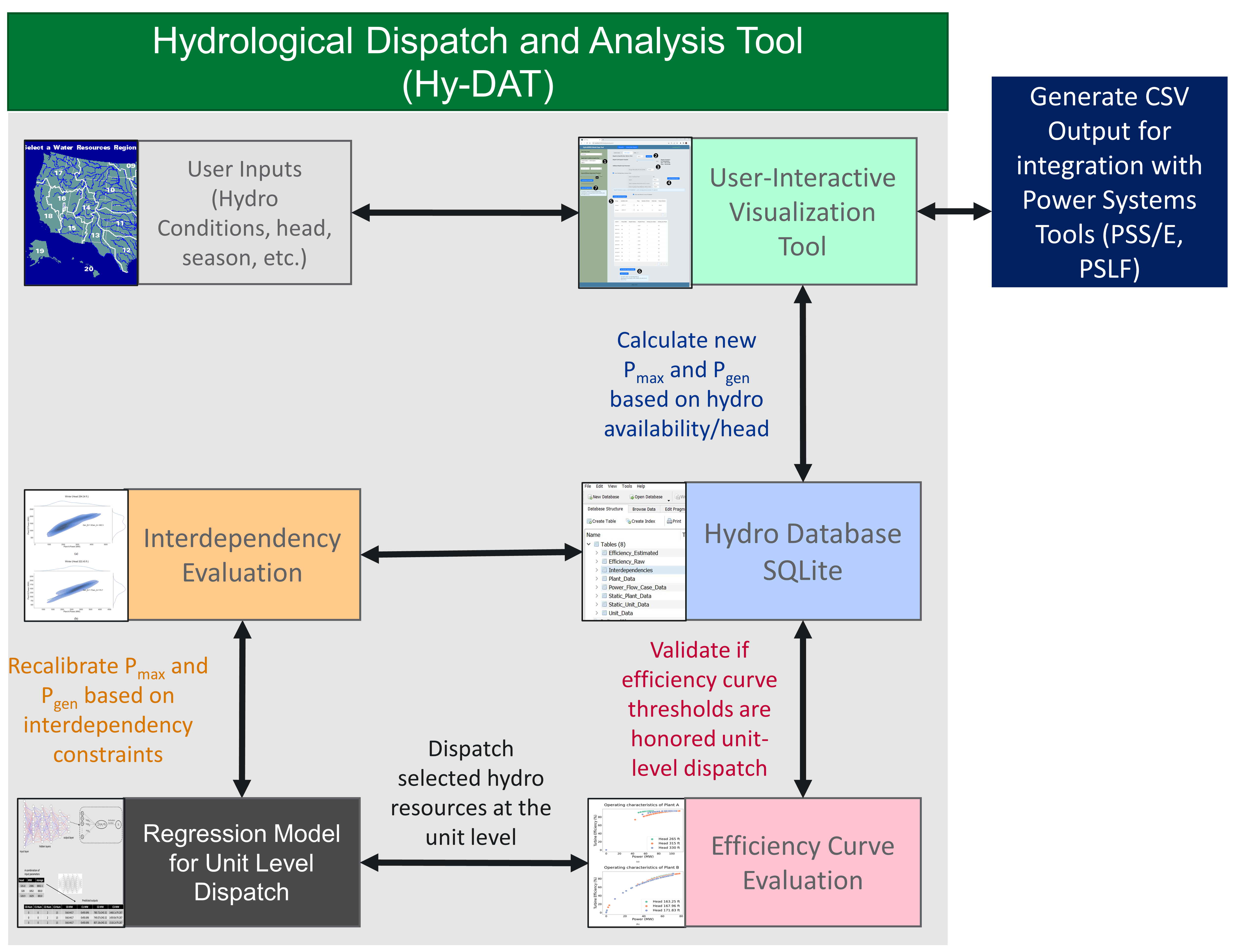}
    \caption{Integration using GUI.}
    \label{fig:Integration}
\end{figure}

To integrate the individual tools and algorithms described above, a GUI is developed for the Hy-DAT tool as an interface for the hydropower database and the suite of tools as illustrated in Fig. \ref{fig:Integration}.

\begin{table}[]
\centering
\caption{Unit-level Dispatch Results}
\label{lev_dis}
\begin{tabular}{|c|c|c|c|c|c|c|}
\hline
Project & \begin{tabular}[c]{@{}c@{}}Unit \\ ID\end{tabular} & \begin{tabular}[c]{@{}c@{}}$P_{gen}$\\ (MW)\end{tabular} & \begin{tabular}[c]{@{}c@{}}$P_{max}$\\ (MW)\end{tabular} & \begin{tabular}[c]{@{}c@{}}Head\\ (ft.)\end{tabular} & \begin{tabular}[c]{@{}c@{}}$P_{gen}$ \\ calculated \\ (MW)\end{tabular} & \begin{tabular}[c]{@{}c@{}}$P_{max}$ \\ available\\ (MW)\end{tabular} \\ \hline
Plant A & 1 & 600 & 707 & 307.1 & 361 & 513.65 \\ \hline
Plant A & 1 & 650 & 707 & 307.1 & 361 & 513.65 \\ \hline
Plant A & 1 & 650 & 707 & 307.1 & 361 & 513.65 \\ \hline
Plant A & 1 & 614.2 & 825.7 & 307.1 & 0 & 599.88 \\ \hline
Plant A & 1 & 700 & 825.7 & 307.1 & 553.97 & 599.88 \\ \hline
Plant A & 1 & 700 & 825.7 & 307.1 & 553.97 & 599.88 \\ \hline
Plant A & 1 & 105 & 125 & 307.1 & 79.48 & 90.81 \\ \hline
Plant A & 2 & 105 & 125 & 307.1 & 79.48 & 90.81 \\ \hline
Plant A & 3 & 105 & 125 & 307.1 & 79.48 & 90.81 \\ \hline
\end{tabular}
\end{table}

 The GUI allows users to select specific or all hydropower plants and specify a hydro condition, either using a historical timeframe or by providing specific hydropower conditions like dry/wet/average water year and season of operation. Based on user inputs, a visualization is also available to plot and evaluate the various hydrological parameters for historical years. From these user inputs, the tool calculates the MW target and available capacity for the selected hydropower resources. However, due to interdependency constraints, some plants downstream, especially the run-of-river plants, have to recalibrate the MW target based on water availability and release from the plants/dams upstream. The recalibrated MW targets are passed on the unit-level dispatch tool, which uses trained artificial intelligence regression models generated from the historical database to dispatch selected resources based on the available head, water availability, and MW target for the plant. Finally, the unit-level dispatches are validated using precalculated efficiency curve datasets from the database to make sure units are dispatched above the efficiency threshold or the unit dispatches are corrected. Finally, the updated MW target and hydropower operational characteristics are available to be exported for planning and operational studies, as shown in Table \ref{lev_dis}. For each generator, it contains unit information, power output from an existing power flow case for reference, and dispatch power output. 

\vspace{-1em}

\section{Concluding Remarks} \label{sec:con}

Operation and planning reliability studies currently do not consider environmental conditions and constraints such as water availability and interdependencies among power plants. The reason for such an approach is lack of coupling of hydro conditions and constraint with the electrical model used in power system planning and operation studies (power flow and dynamic models). Industry develops base cases for heavy winter and summer loading conditions as well as shoulder cases representing light loading conditions. In the environment of large renewable penetration of intermittent resources (solar and wind) and carbon-free goals, hydro becomes a more critical resource. For this reason, it is important to properly represent hydro conditions in base cases and, similar to developing cases having heavy and light load conditions, we need to have high and low hydro conditions properly represented in steady-state and dynamic models used for power system planning and operation reliability studies. This paper provides an end-to-end methodology for addressing modeling gaps in hydropower resources using Hy-DAT, a newly developed tool that incorporates a database and set of tools and algorithms to use historical data in accurately dispatching hydropower resources. 


\vspace{-1em}
\bibliographystyle{IEEEtran}
\bibliography{references.bib}

\begin{thebibliography}{10}
\providecommand{\url}[1]{#1}
\csname url@samestyle\endcsname
\providecommand{\newblock}{\relax}
\providecommand{\bibinfo}[2]{#2}
\providecommand{\BIBentrySTDinterwordspacing}{\spaceskip=0pt\relax}
\providecommand{\BIBentryALTinterwordstretchfactor}{4}
\providecommand{\BIBentryALTinterwordspacing}{\spaceskip=\fontdimen2\font plus
\BIBentryALTinterwordstretchfactor\fontdimen3\font minus
  \fontdimen4\font\relax}
\providecommand{\BIBforeignlanguage}[2]{{%
\expandafter\ifx\csname l@#1\endcsname\relax
\typeout{** WARNING: IEEEtran.bst: No hyphenation pattern has been}%
\typeout{** loaded for the language `#1'. Using the pattern for}%
\typeout{** the default language instead.}%
\else
\language=\csname l@#1\endcsname
\fi
#2}}
\providecommand{\BIBdecl}{\relax}
\BIBdecl

\bibitem{mitra2023gaps}
B.~Mitra, S.~Datta, S.~Kincic, N.~Samaan, and A.~Somani, ``Gaps in
  representations of hydropower generation in steady-state and dynamic
  models,'' in \emph{2024 ISGT NA}, 2024.

\bibitem{kincic2023hydropower}
S.~Kincic, N.~A. Samaan, S.~Datta, A.~Somani, J.~Tan, T.~Mosier, R.~Bhattarai,
  and H.~Yuan, ``Hydropower modeling gaps in planning and operational
  studies,'' Pacific Northwest National Lab.(PNNL), Richland, WA (United
  States), Tech. Rep. PNNL-33836, 2023.

\bibitem{kosterev}
D.~Kosterev, ``Hydro turbine-governor model validation in pacific northwest,''
  \emph{IEEE Transactions on Power Systems}, vol.~19, no.~2, 2004.

\bibitem{USACE}
\BIBentryALTinterwordspacing
U.~A. C.~O. ENGINEERS, ``Usace.'' [Online]. Available:
  \url{https://www.nwd-wc.usace.army.mil/dd/common/dataquery/www/}
\BIBentrySTDinterwordspacing

\bibitem{Britishhydro}
\BIBentryALTinterwordspacing
U.~B.~H. Association, ``A guide to uk mini-hydro development,'' 2005. [Online].
  Available: \url{https://www.british-hydro.org/micro-hydro-guide/}
\BIBentrySTDinterwordspacing

\bibitem{DB}
\BIBentryALTinterwordspacing
M.~Piacentini, ``Db browser.'' [Online]. Available:
  \url{https://github.com/datacarpentry/sql-socialsci/blob/gh-pages/CITATION}
\BIBentrySTDinterwordspacing

\bibitem{TSAT}
C.~Powertech Labs Inc.~Surrey, British~Columbia, ``Transient security
  assessment tool user manual,'' 2011.

\bibitem{USACEMIL}
U.~B.~H. Association, ``Us army corps of engineers.''

\bibitem{BelgiumHydro}
E.~S.~H. Association, ``Guide on how to develop a small hydropower plant,''
  2004.

\bibitem{BHATTI2023120894}
B.~A. Bhatti, S.~Hanif, J.~Alam, B.~Mitra, R.~Kini, and D.~Wu, ``Using energy
  storage systems to extend the life of hydropower plants,'' \emph{Applied
  Energy}, vol. 337, p. 120894, 2023.

\bibitem{Bhaskar_hydrogenerate}
\BIBentryALTinterwordspacing
B.~Mitra, J.~F. Gallego-Calderon, S.~N. Elliott, T.~M. Mosier, C.~J.
  Bastidas~Pacheco, U.~O. of~Energy~Efficiency, and R.~Energy, ``Hydrogenerate:
  Open source python tool to estimate hydropower generation time-series,
  version 3.6 or newer,'' 10 2021. [Online]. Available:
  \url{https://www.osti.gov//servlets/purl/1829986}
\BIBentrySTDinterwordspacing

\bibitem{mitra2020development}
\BIBentryALTinterwordspacing
B.~Mitra, S.~Elliott, J.~Koudelka, C.~Davis, T.~Mosier, D.~Anderson, and
  J.~Saulsbury, ``Development and role of microgrids surrounding modernized
  irrigation district,'' in \emph{AGU Fall Meeting Abstracts}, 2020. [Online].
  Available: \url{https://irrigationviz.pnnl.gov/docs/modules/hydropower}
\BIBentrySTDinterwordspacing

\bibitem{columbia}
\BIBentryALTinterwordspacing
T.~C. R. S.~I. Story, ``Second edition, bonneville power administration
  (bpa),'' April 2001. [Online]. Available:
  \url{http://www.bpa.gov/power/pg/columbia\_river\_inside\_story.pdf}
\BIBentrySTDinterwordspacing

\bibitem{wang2021machine}
D.~Wang, J.~Bao, Z.~Xu, B.~Koeppel, O.~A. Marina, A.~Noring,
  M.~Zamarripa-Perez, A.~Iyengar, E.~Eggleton, D.~T. Schwartz \emph{et~al.},
  ``Machine learning tools set for natural gas fuel cell system design,''
  \emph{ECS Transactions}, vol. 103, no.~1, p. 2283, 2021.

\end{thebibliography}

\end{document}